\documentclass[twocolumn,showpacs,preprintnumbers,amsmath,amssymb,prl,aps,10pt]{revtex4-1}

\usepackage{graphicx}
\usepackage{dcolumn}
\usepackage{bm}

\usepackage{mathrsfs,amsbsy,amssymb,latexsym,amsfonts,amsmath}
\usepackage{graphicx,color}

\newcommand{\dd}{{\rm d}}

\newcommand{\tf}{T_{\rm F} }

\newcommand{\sym}{{\rm SYM} }

\newcommand{\2}{{\it 2} }

\begin{document}

\title{The Schwinger pair production rate in confining theories via holography}

\author{Daisuke Kawai}
\author{Yoshiki Sato}
\author{Kentaroh Yoshida}
\affiliation{Department of Physics, Kyoto University, 
Kyoto 606-8502, Japan}

\begin{abstract}
We study the Schwinger pair production in confining theories. 
The production rate in an external electric field $E$ 
is numerically evaluated by using the holographic description. There exist two kinds of 
critical values of the electric field: (i) $E=E_{\rm c}$\,,  
above which there is no potential barrier and particles are freely generated, and
(ii) $E=E_{\rm s}$\,, below which the confining string tension dominates the electric field 
and the pair production does not occur. 
We argue the universal exponents associated with the critical behaviors.  
\end{abstract}

\date{\today}


\maketitle


In quantum electrodynamics (QED) vacuum, virtual pairs of particle and antiparticle are 
momentarily created and annihilated. In the presence of a strong electric field, the virtual pair 
can become real particles. This phenomenon is known as the Schwinger effect \cite{Schwinger}. 
The production rate is evaluated under the weak field condition in a weakly coupled region \cite{Schwinger}. 
It is generalized to an arbitrary coupling \cite{AAM}. 

The Schwinger effect is not intrinsic to QED, and it is ubiquitous in quantum field theories, 
including the fundamental matter fields in the presence of a strong electric field. It can also be 
argued in the context of the AdS/CFT correspondence \cite{M,GKP,W}, 
where AdS and CFT are anti-de Sitter space and conformal field theory, respectively.   
By Higgsing the planar $\mathcal{N}=4$ $SU(N+1)$ super Yang-Mills (SYM) theory, 
the production rate of the fundamental particles is evaluated. 
The resulting action is composed of the three parts: 
\[
S^{SU(N+1)}_{\mathcal{N}=4 \,\sym} = S^{SU(N)}_{\mathcal{N}=4 \,\sym} + S^{U(1)}_{\mathcal{N}=4 \,\sym} + S_{\rm W}\,.
\]
Here the key ingredient is the action $S_{\rm W}$ of the fundamental fields with the covariant derivative 
\[
D_{\mu} = \partial_{\mu} + i a_{\mu} -i A_{\mu}\,,  
\]
where $a_{\mu}$ and $A_{\mu}$ are $U(1)$ and $SU(N)$ gauge fields, respectively. 

In the large $N$ limit, the scheme of \cite{AAM} is applicable
by taking $a_{\mu}$ as a source of an external electric field and $A_{\mu}$ as a dynamical field, 
where the fluctuation of $a_{\mu}$ can be ignored in this limit. However, this approach encounters 
a puzzle of the critical electric field, above which 
the production rate is not exponentially suppressed anymore. 
The phenomenon of this kind occurs also in string theory \cite{max1,max2}. 
The critical value obtained from the production rate disagrees with the one 
derived from the Dirac-Born-Infeld (DBI) action of a probe D3-brane in the bulk ${\rm AdS}_5$. 

Semenoff and Zarembo solved this puzzle by considering a probe D3-brane at the intermediate position 
in the bulk ${\rm AdS}_5$ \cite{SZ}.  
Then the production rate $P$ (per unit time and volume) is evaluated by computing  
the expectation value of a circular Wilson loop on the probe D3-brane 
in the holographic description with the Nambu-Goto (NG) action coupled 
to a constant electric NS-NS 2-form $B_\2 = B_{01}\dd x^0\wedge \dd x^1$\,, 
where NS is an abbreviation for Neveu-Schwarz. 
Then $P$ is evaluated as 
\begin{eqnarray}
P &\sim& \exp(-S_{\rm NG} -S_{B_{\it 2}})  \nonumber \\ 
 &=& \exp \left[-\frac{\sqrt{\lambda}}{2} 
\left(\sqrt{\frac{E_{\rm c}}{E}} - \sqrt{\frac{E}{E_{\rm c}}}\right)^2\right]\,, \label{SZ}
\end{eqnarray}
where $\lambda \equiv Ng_{\rm YM}^2$ is the 't~Hooft coupling and $E$ is an electric field. 
The critical electric field $E_{\rm c}$ is  
\begin{eqnarray}
E_{\rm c} = \frac{2\pi m^2}{\sqrt{\lambda}} \label{Ec}
\end{eqnarray}
and agrees with the DBI result. 
This prescription has been generalized to various cases \cite{BKR,SY}. 
From the viewpoint of the potential analysis, one encounters another puzzle 
if the usual Coulomb potential \cite{Wilson1,Wilson2} is utilized, as raised in \cite{SZ}. 
This puzzle has been solved by using a modified Coulomb potential, and now 
the agreement of the critical electric field is supported also by the potential analysis \cite{SY2}. 

\begin{figure}[htbp]
\vspace*{-0.2cm}
\begin{center}
\includegraphics[scale=.4]{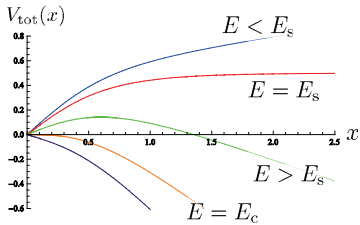}
\end{center}
\vspace*{-0.7cm}
\caption{The total potential in a confining theory on a D3-brane with a compact circle \cite{SY3}. \label{potential:fig}} 
\end{figure}

An intriguing subject in the holographic computation is to study the Schwinger effect 
in confining gauge theories. It is quite difficult to study it analytically, and the lattice formulation is not 
applicable to computing the pair production rate. However, the holographic potential 
analysis is so powerful as to show the total potential numerically as in FIG.\,\ref{potential:fig} 
even in confining theories. Then there is another critical value $E=E_{\rm s}$, 
below which the Schwinger effect cannot occur. When the electric field is not stronger 
than the confining string tension ($E<E_{\rm s}$), the fundamental particles are still confined. However, 
when $E>E_{\rm s}$\,, the particles may be liberated via the Schwinger effect. 
This phenomenon is an analogue of the electrical breakdown 
of insulators in condensed matter physics, e.g., one-dimensional Hubbard models \cite{OA}. The similar behavior 
is also shown in lattice gauge theories \cite{Yamamoto}.

An important issue is to compute the pair production rate in the confining gauge theories 
by following the prescription in \cite{SZ}. It seems likely difficult to compute 
{\it analytically} the area of the classical string solution corresponding to a circular Wilson loop. 
However, it can be evaluated numerically and the result exhibits the expected behavior 
from the potential analysis. 


\medskip

{\it Setup}.---There are various gravitational solutions dual to confining gauge theories. 
Here we concentrate on an AdS soliton background composed of D3-branes \cite{Witten,HM} 
\begin{align}
\dd s^2&=\frac{L^2} {z^2}\left[ -(\dd x^0)^2+\sum _{i=1}^{2}(\dd x^i )^2+
f(z)(\dd x^3 )^2+\frac{\dd z^2}{f(z)}\right]  \notag \\
& + L^2 \dd s^2_{{\rm S}^{5}}\,, \label{metric} 
\qquad f(z)=1-\left( \frac{z}{z_{\rm t}}\right)^{4}\,,
\end{align}
where $L$ is the AdS radius and the $x^3$ direction is compactified on a circle ${\rm S}^1$ 
with radius $R = \pi z_{\rm t}$.  
The confining string tension is proportional to $1/z_{\rm t}^2$ at long distances. 
The internal space is simply assumed to be ${\rm S}^{5}$. 
The AdS boundary is located at $z=0$\,, and the spacetime terminates at $z=z_{\rm t}$. 
A probe D3-brane is put at $z=z_0~( 0<z_0 < z_{\rm t})$, where $z_0$ determines a dynamical scale  
in the gauge-theory language. Then the $SU(N)$ gauge field causes a linear potential between the fundamental 
matter fields.  

\begin{figure}[htbp]
\begin{center}
\includegraphics[scale=.3]{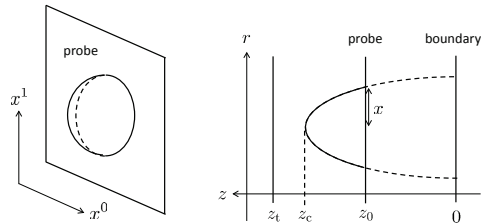}
\end{center}
\vspace*{-0.65cm}
\caption{The configuration of the classical string world sheet.} \label{configuration:fig}
\vspace*{-0.2cm}
\end{figure}

To evaluate the pair production rate, the expectation value 
of a circular Wilson loop has to be computed with the holographic description \cite{Wilson1,Wilson2}. 
In the following, we work in the Euclidean signature. 
For the classical string solution that corresponds to the circular Wilson loop, 
the following ansatz is supposed as usual \cite{BCFM,DGO}, 
\begin{equation}
x^0=r(\sigma)\cos \tau \,, \quad x^1=r(\sigma)\sin \tau \,, \quad z=z(\sigma)\,,
\label{target}
\end{equation}
and the other components are set to be zero.  
The string world-sheet coordinates $(\tau,\sigma)$ are restricted to the following range,  
\[
0 \leq \tau < 2\pi\,, \quad 0 \leq \sigma \leq \sigma_0\,, 
\] 
and the boundary conditions for $r(\sigma)$ and $z(\sigma)$ are supposed as 
\begin{eqnarray}
r(0) =0\,, ~~ r(\sigma_0) = x\,,  ~~ 
z(0) =z_{\rm c}\,, ~~ z(\sigma_0) = z_0\,. \label{bdc}
\end{eqnarray}
Here the shape of the solution is assumed to be cuplike, as depicted in FIG.\,\ref{configuration:fig},
and $x$ describes the radius of the circular Wilson loop on the probe D-brane. 

With the ansatz, the string action is rewritten as 
\begin{eqnarray}
S_{\rm NG} &=& 2\pi L^2 \tf \int_0^x \! \dd r \, \frac{r}{z^2}\sqrt{1+\frac{z'^2}{f(z)}}\,, \\ 
S_{B_{\it 2}} &=& -2\pi \tf B_{01} \int_0^x\!\! \dd r\, r =-\pi E x^2\,,  
\end{eqnarray}
where $T_{\rm F}$ is the string tension and the nonvanishing component of $B_\2$ is $B_{01}=E/T_{\rm F}$. 
Note that the integration variable is converted from $\sigma$ to $r$\,. 
Then the function $z(r)$ satisfies the ordinary differential equation,  
\begin{equation}
z'+\frac{2r f(z)}{z} + r z'' - \frac{r z'^2}{2f(z)} \frac{\dd f}{\dd z}(z)   +\frac{z'^3}{f(z)}+\frac{2r z'^2}{z}=0\,. 
\label{diff}
\end{equation}
The remaining task is to solve (\ref{diff}) with the boundary conditions (\ref{bdc}). 
It seems quite difficult to perform it analytically, but it is possible numerically. 

Before going to the numerical analysis, one more step is needed,
which is to take a variation with respect to $x$ in the prescription 
of \cite{SZ}. This would make the numerical analysis much harder. However, 
this step can be replaced by imposing an additional condition to 
the classical string solution as pointed out in \cite{SY}. 
This is the mixed boundary condition for the string coordinates in the presence of $B_\2$ and leads to 
the constraint condition,  
\begin{equation}
z'=-\sqrt{f(z)\left( \frac{1}{\alpha^2}-1\right)} \,\Biggr|_{z=z_0}\,, 
\label{mixed}
\end{equation}
where $\alpha$ is defined as  
\begin{eqnarray}
\alpha \equiv \frac{E}{E_{\rm c}}\,, \quad E_{\rm c} \equiv \tf \frac{L^2}{z_0^2}\,.  
\end{eqnarray}
The resulting classical action depends on $E$\,, through this constraint.  


\medskip 

\begin{figure}[htbp]
\vspace*{-0.6cm}
\[
\begin{array}{cc}
\includegraphics[scale=.19]{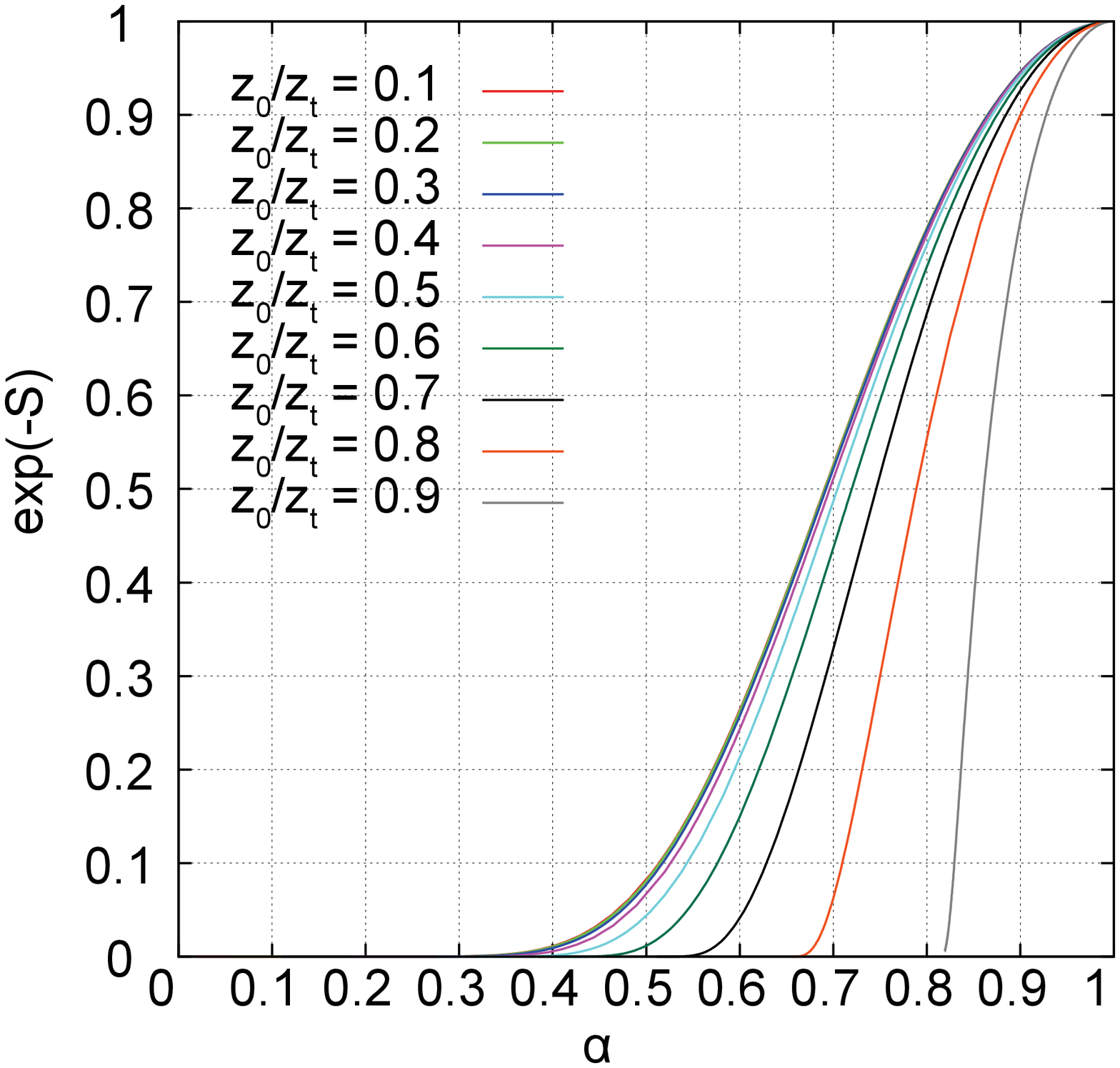} &
\includegraphics[scale=.19]{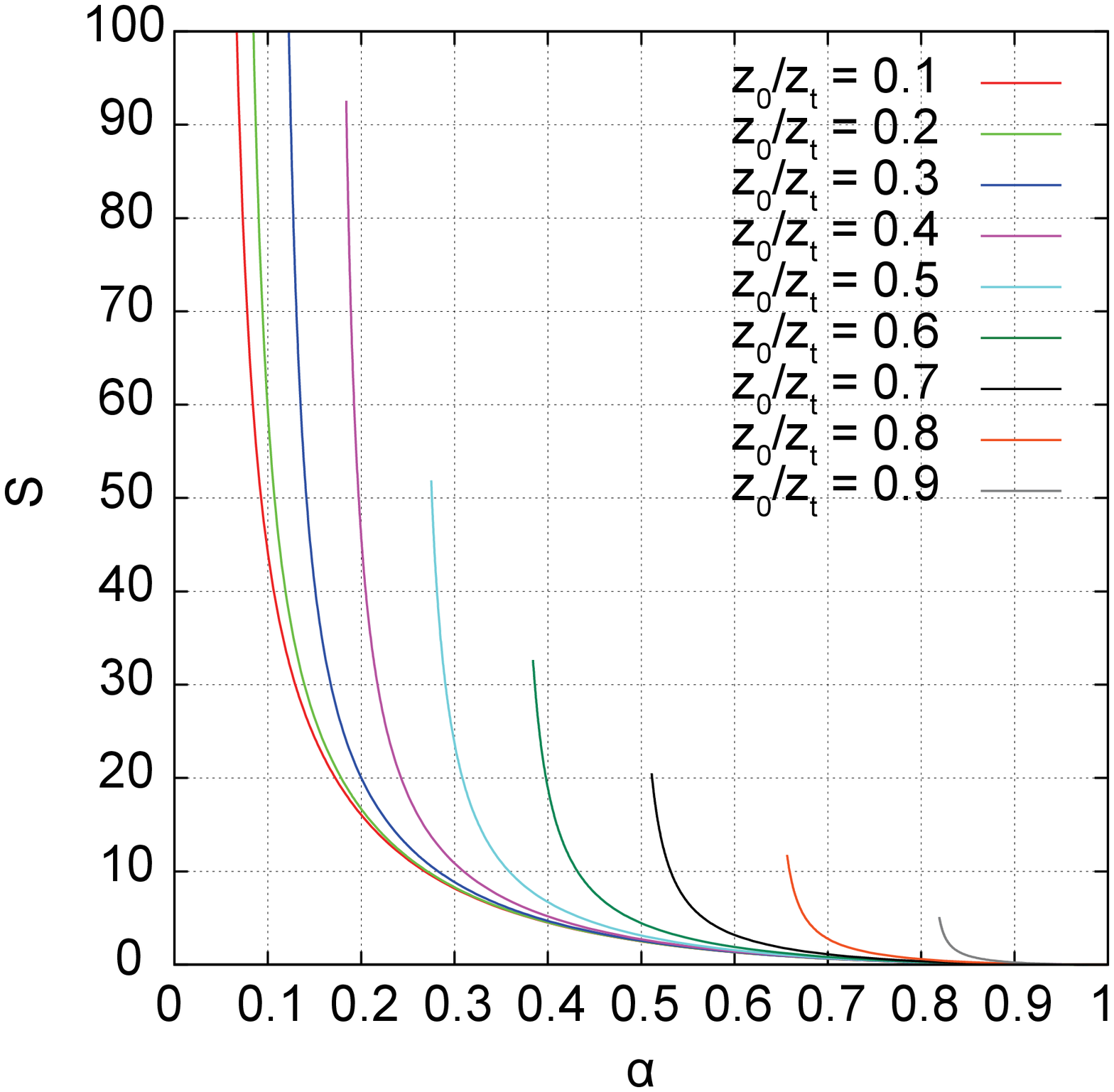} \\ 
\vspace*{0.3cm} \mbox{\footnotesize (a) \quad the exp factor} & 
\mbox{\footnotesize (b) \quad the classical action}
\end{array}
\]
\vspace{-0.7cm}
\caption{The plots of the exp factor and the classical action.} 
\label{rate:fig}
\end{figure}

{\it Numerical results}.---The next is to evaluate numerically $z(r)$ satisfying the differential equation (\ref{diff}) and the conditions 
(\ref{bdc}) and (\ref{mixed}). Up to technical difficulties, the strategy of the numerical analysis is straightforward. 
The exponential part and the classical action can be evaluated numerically, 
and the results are shown in FIG.\,\ref{rate:fig}. Here the parameters are taken as $2\pi T_{\rm F}L^2 =10$ 
(i.e., $\lambda=100$) for the supergravity approximation.  
The ratio $z_0/z_{\rm t}$ is the inverse of the compactification radius measured 
by the dynamical scale of the system (up to $1/\pi$).  

As shown in FIG.\,\ref{rate:fig} (a), the exponential factor almost vanishes below a certain value of $E$\,. 
 In other words, the classical action diverges around the value as shown in FIG.\,\ref{rate:fig} (b). 
This numerical value should be compared with the value of $E_{\rm s}$ defined as 
\begin{eqnarray}
E_{\rm s} \equiv \tf \frac{L^2}{z_{\rm t}^2}\,, 
\end{eqnarray}
which has been derived from the potential analysis at long distances \cite{SY3}. 
Although the plots in FIG.\,\ref{rate:fig} (a) have a long tail, it seems that the value of $E_{\rm s}$ agrees with 
the threshold value in FIG.\,\ref{rate:fig}. 

The results in FIG.\,\ref{rate:fig} are further supported 
by the asymptotic behaviors around $E=E_{\rm s}$ and $E=E_{\rm c}$\,. 

\medskip 


{\it The behavior around $E=E_{\rm s}$}.---Let us estimate the behavior of 
the classical action $S$ around $E = E_{\rm s}$\,. When $E < E_{\rm s}$, 
the Schwinger effect does not occur, as indicated by the potential analysis \cite{SY3}. 
Hence, $S$ should diverge at $E=E_{\rm s}$\,, and the leading term is given by  
\begin{eqnarray}
S = \frac{f(\alpha,\alpha_{\rm s})}{(\alpha - \alpha_{\rm s})^{\gamma_{\rm s}}} + \cdots\,, 
\quad \alpha_{\rm s} \equiv \frac{E_{\rm s}}{E_{\rm c}} = \left(\frac{z_{0}}{z_{\rm t}}\right)^2\,, 
\end{eqnarray}
where $\gamma_{\rm s}$ is a positive exponent that may depend on $\alpha_{\rm s}$
and $f(\alpha,\alpha_{\rm s})$ is a regular function at $E=E_{\rm s}$. 

\begin{figure}[htbp]
\vspace*{-0.5cm}
\[
\begin{array}{cc}
\includegraphics[scale=.19]{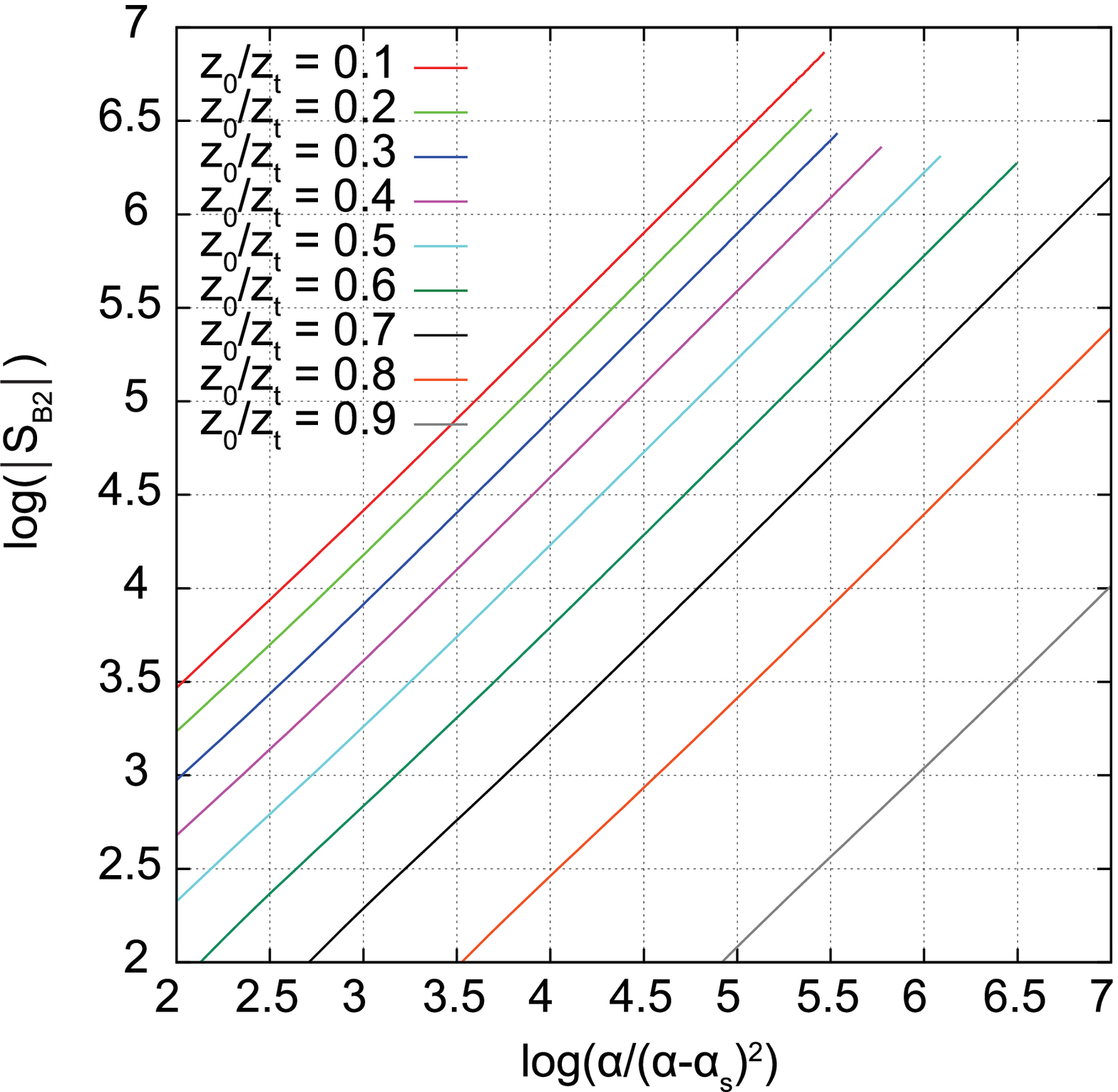} & 
\includegraphics[scale=.18]{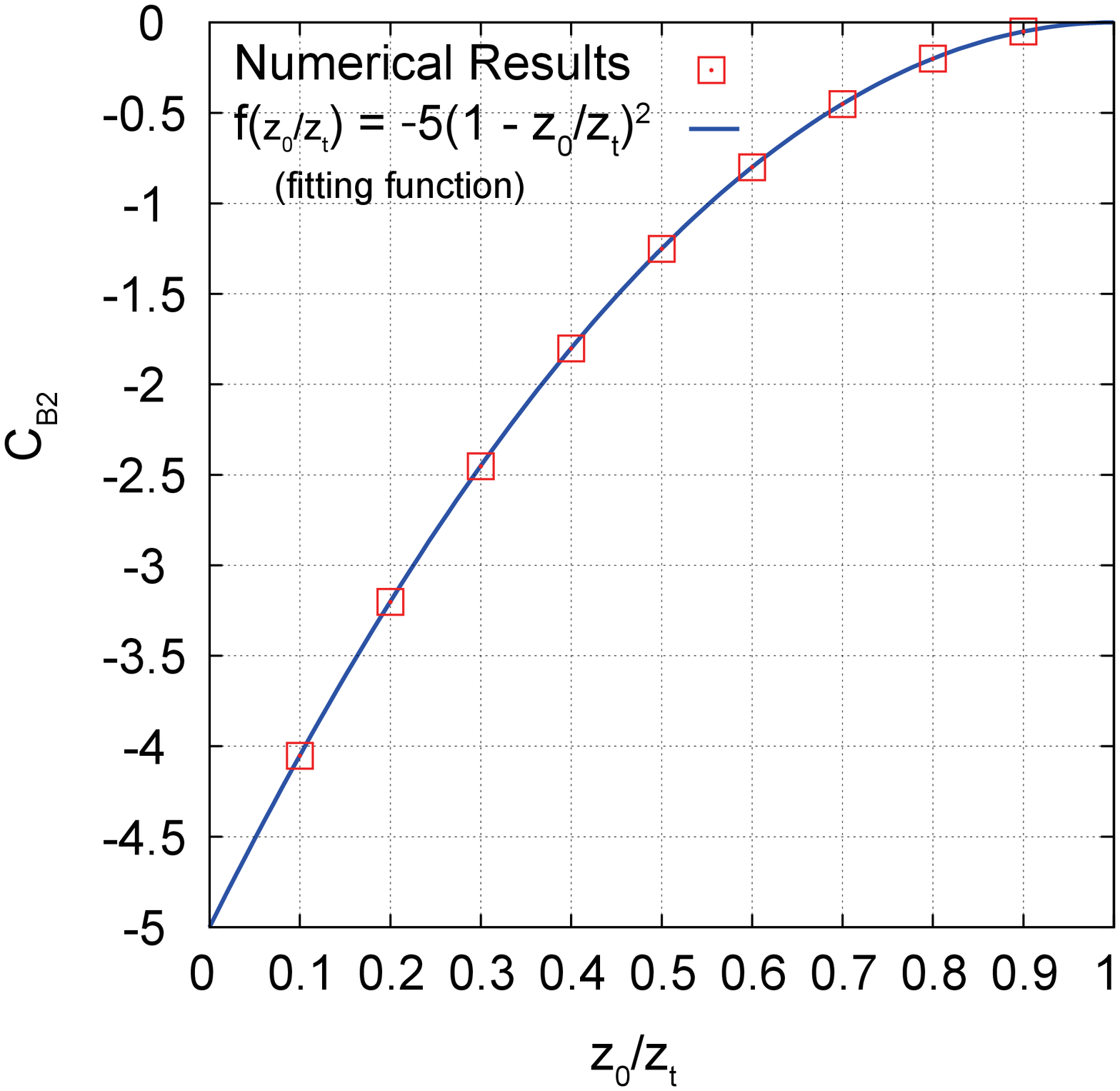} 
\vspace*{0.1cm} 
\\ 
\mbox{\footnotesize (a) \quad $S_{B_\2}$} & 
\mbox{\footnotesize (b) \quad $C_{B_\2}(\alpha_{\rm s})$} 
\end{array}
\]
\vspace*{-0.5cm}
\caption{\footnotesize The behavior of $S_{B_\2}$ and $C_{B_\2}(\alpha_{\rm s})$ around $E=E_{\rm s}$. 
\label{B2:fig}}\vspace*{-0.5cm}
\end{figure}

It is convenient to divide $S$ into the NG part $S_{\rm NG}$ and the $B_\2$ part $S_{B_\2}$. 
The behavior of $S_{B_\2}$ is relatively simple, and it behaves as 
\begin{eqnarray}
S_{B_\2} = \frac{C_{B_\2}(\alpha_{\rm s})\alpha}{(\alpha-\alpha_{\rm s})^2} + \mbox{the regular}. 
\end{eqnarray}
The plots in FIG.\ref{B2:fig} confirm this form, and 
$C_{B_\2}(\alpha_{\rm s})$ is determined from the numerical data like 
\begin{eqnarray}
C_{B_\2}(\alpha_{\rm s}) = -\frac{\sqrt{\lambda}}{2}\left(1- \sqrt{\alpha_{\rm s}}\,\right)^2\,. 
\end{eqnarray}

\begin{figure}[htbp]
\vspace*{-0.6cm}
\[
\begin{array}{cc}
\includegraphics[scale=.18]{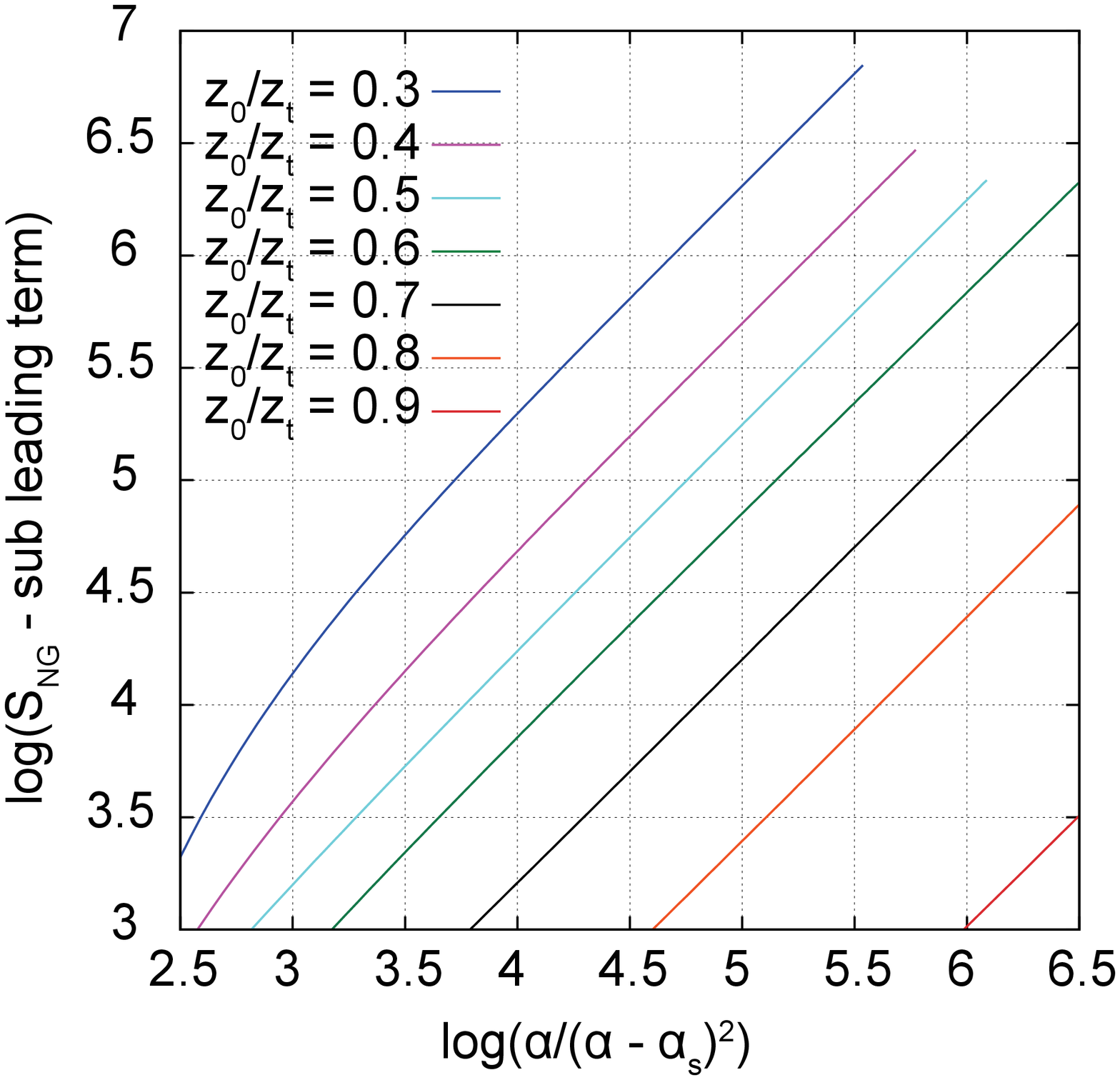} & 
\includegraphics[scale=.18]{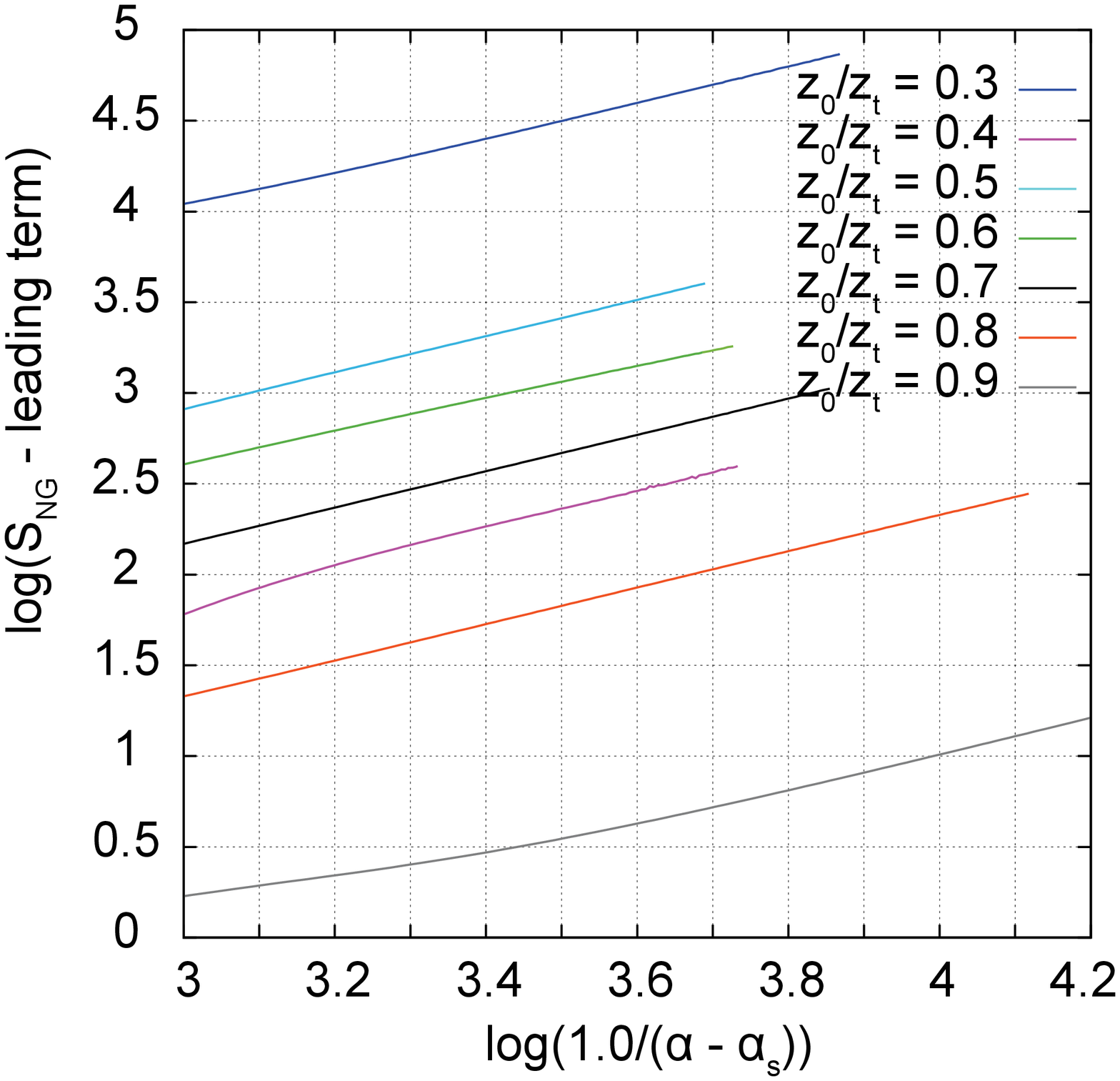} 
\\ 
\mbox{\footnotesize (a)~~the leading term} & 
\mbox{\footnotesize (b)~~the sub-leading term} 
\end{array}
\]
\vspace*{-0.5cm}
\caption{\footnotesize The leading behavior of $S_{\rm NG}$ around $E=E_{\rm s}$.  \label{NG:fig}}
\end{figure}

For the NG part, suppose the following ansatz:
\begin{eqnarray}
S_{\rm NG} = \frac{C_{\rm NG}(\alpha_{\rm s})\alpha}{(\alpha-\alpha_{\rm s})^2} 
+ \frac{D_{\rm NG}(\alpha_{\rm s})}{\alpha-\alpha_{\rm s}} 
+ \mbox{the regular}. 
\end{eqnarray} 
This ansatz is also supported from the numerical analysis. The leading term and the subleading term 
are shown in FIG.\,\ref{NG:fig}. Each of them is determined with the help of the method of least squares. 

Note here that the plots in FIG.\,\ref{NG:fig} are valid, roughly for $\alpha_{\rm s} \gtrsim 0.1$ 
or equivalently $z_0/z_{\rm_t} \gtrsim 0.32$\,. This comes from the limitation of numerical precision. 
For example, if we want to consider $\alpha_{\rm s}=0.01$,  then $\alpha$ can approach to $\alpha_{\rm s}$ 
at most as $\alpha=0.02$ in the present analysis. 
Then $\alpha/(\alpha-\alpha_{\rm s})^2 = 2 \times 10^2$ and  $1/(\alpha-\alpha_{\rm s}) = 10^2$\,. 
Thus, it seems difficult to distinguish the leading term from the subleading one. To make matters worse, 
more dominant parts might enter sneakily in our analysis. Much higher accuracy is needed for $\alpha$ 
so as to eliminate the undesirable contributions. To consider $\alpha_{\rm s}=0.01$ properly, 
the numerical precision like $\alpha=0.01005$ is typically necessary. 
This accuracy implies that the graphs in FIG.\ref{NG:fig} 
have to be much more extended to the right. 

Thus, in total, the classical action behaves as 
\begin{equation}
S = \frac{C(\alpha_{\rm s})\,\alpha }{(\alpha -\alpha _{\rm s})^2}+\frac{D(\alpha_{\rm s})}{\alpha -\alpha _{\rm s}} 
+ \mbox{the regular}\,,  
\end{equation}
where the following quantities are introduced, 
\[
C(\alpha_{\rm s}) \equiv C_{\rm NG}(\alpha_{\rm s}) + C_{B_\2}(\alpha_{\rm s})\,, \quad 
D(\alpha_{\rm s}) \equiv D_{\rm NG}(\alpha_{\rm s})\,. 
\]
Note that the exponent of the leading term, 
\[
\gamma_{\rm s}=2\,, 
\] 
and that of the subleading term are universally fixed for $\alpha_{\rm s} \gtrsim 0.1$. 
Note that this behavior is reliable for $\alpha_{\rm s} \lesssim 0.4$. 
The absolute value of $C(\alpha_{\rm s})$ becomes almost zero for $\alpha_{\rm s} \gtrsim 0.4$, 
but the vanishing $C({\alpha_{\rm s}})$ cannot be definitely stated due to the precision of 
the numerical method. It may suggest the existence of a phase transition.  


\medskip 

\begin{figure}[htbp]
\vspace{-0.8cm}
\[
\begin{array}{cc}
\includegraphics[scale=.18]{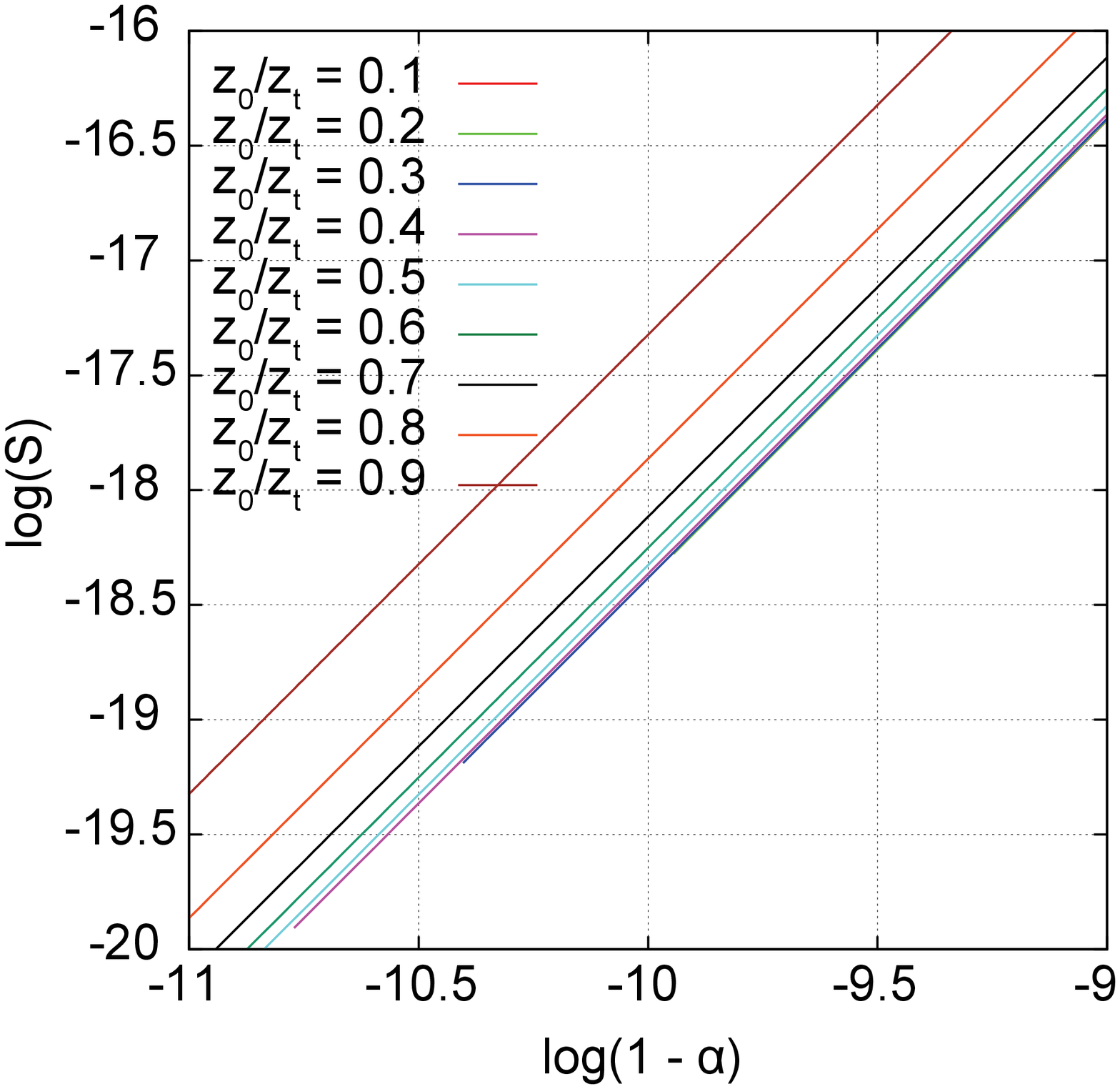} & 
\includegraphics[scale=.17]{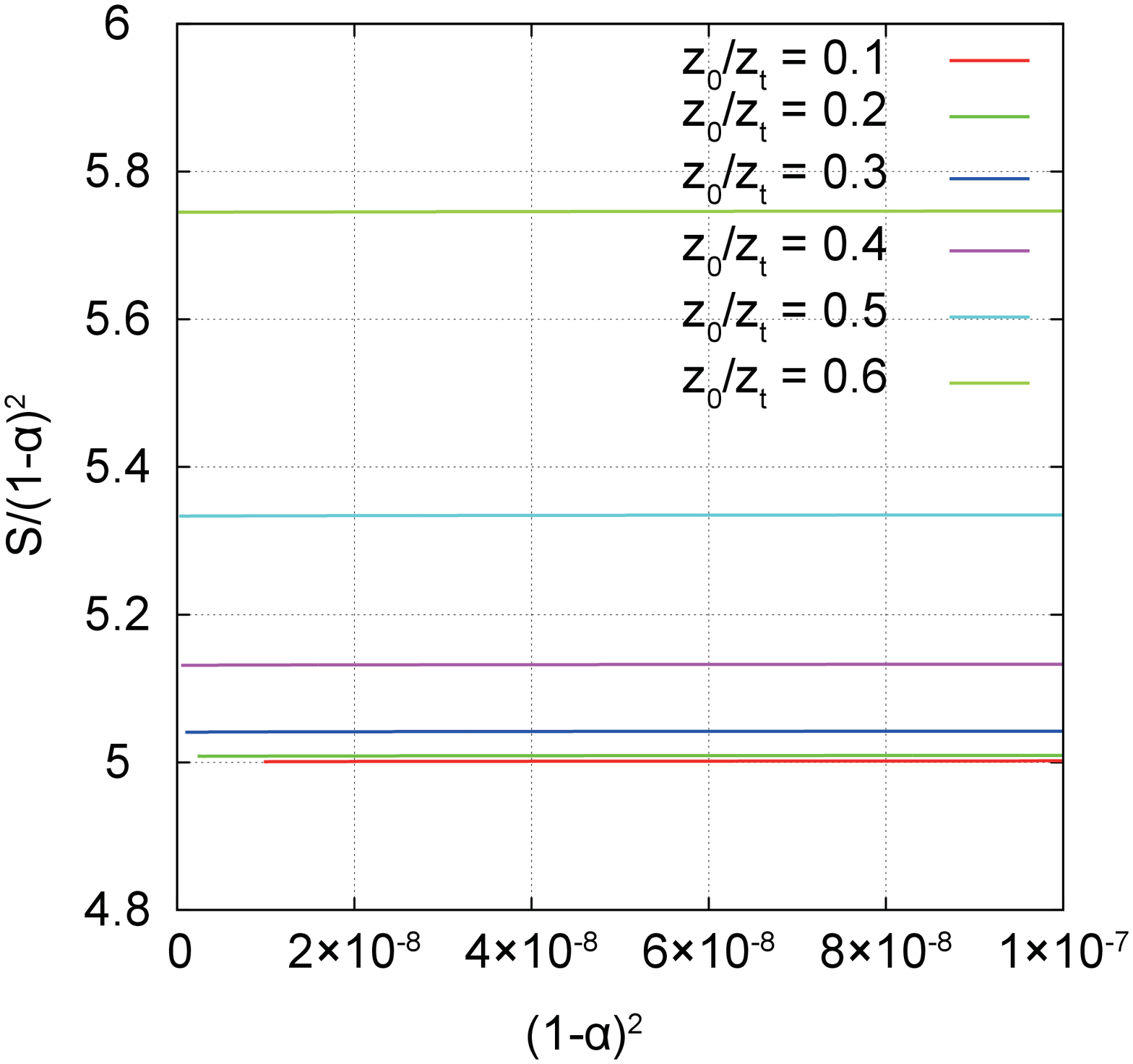} 
\vspace{-0.1cm}\\ 
\mbox{\footnotesize (a) the classical action } & \mbox{\footnotesize (b) ~~$B(\alpha_{\rm s})$}
\end{array}
\]
\vspace*{-0.6cm}
\caption{The behavior of the classical action near $E=E_{\rm c}$. 
The log-log plot indicates that $\gamma_{\rm c} =2$ universally. 
The coefficient approaches 5 as $\alpha_{\rm s} \to 0$ because of $\lambda=100$.} 
\label{near-critical:fig}
\vspace*{-0.2cm}
\end{figure}

{\it The behavior near $E=E_{\rm c}$}.---The behavior of $S$  
near the critical value $E_{\rm c}$ is expected as 
\begin{equation}
S = B(\alpha_{\rm s})(1-\alpha)^{\gamma_{\rm c}} + \cdots
\end{equation}
because it vanishes when $\alpha=1$\,, as shown in Fig.\,\ref{rate:fig}. 
Here $B(\alpha_{\rm s})$ is a regular function of $\alpha_{\rm s}$, and 
$\gamma_{\rm c}$ is an exponent. 
The plots in FIG.\,\ref{near-critical:fig} indicate the universal value, 
\[
\gamma_{\rm c} =2\,.
\] 
When $\alpha_{\rm s} =0$, the asymptotic form is obtained from (\ref{SZ}),   
\[
S = \frac{\sqrt{\lambda}}{2}(1-\alpha)^2 + \mathcal{O}\bigl((1-\alpha)^3\bigr)\,;
\]
hence, the universality of $\gamma_{\rm c}$ is also supported from (\ref{SZ}). 
Although we are confined to a confining D3-brane background here, we argue that 
$\gamma_{\rm c}=2$ would be universal for arbitrary confining backgrounds specified 
by the theorem \cite{Son1}. In fact, the universality of the existence of $E_{\rm s}$ and $E_{\rm c}$ 
is shown in \cite{SY4}. 

Finally, it is worth noting on the behavior of 
the classical solution. The boundary circular loop is infinitely huge for $E \leq E_{\rm s}$\,, 
and it gets a finite radius for $E >E_{\rm s}$\,. Then the loop shrinks to zero as $E \to E_{\rm c}$\,. 


\medskip 

{\it Summary and Discussion}.---We have computed the pair production rate of the fundamental particles 
in confining theories realized in a D3-brane background with a compactified spatial direction. 
The production rate of the fundamental particles has been evaluated numerically. 
The result supports quantitatively the behavior indicated by the potential analysis.

There are two kinds of critical behaviors around (1) $E=E_{\rm s}$ and (2) $E=E_{\rm c}$\,. 
The system may be extremely simplified around these values, and 
some universal quantities may be figured out. 
Indeed, the critical exponents have been computed here. 
It is important to check the universality of them for various backgrounds \cite{Son1}. 
The same result is obtained for the D4-brane case \cite{future}. 
The gauge-theory analysis has not been done so far, and 
the mathematical foundation for the universality is not clear. 
The holographic approach would be a good compass to explore it.

Supposing that the universality holds, our result makes a nontrivial prediction for 
QCD in a strong electric field as well as electrical breakdown in insulators. 
The physical observable is the production rate of hadron jets in QCD  or $e^+$-$e^-$ pairs in insulators. 
A standard experiment is to measure the persistence time of the vacuum.  
The predicted exponents may be observed in tabletop experiments.

The Schwinger effect in the confining phase is interpreted as a kind of deconfinement phase transition, 
and it may generalize the QCD phase diagram. 
Due to the presence of an external electric field, the system cannot exhibit the equilibrium state, 
but a nonequilibrium stationary state may be realized. To reveal the feature of the deconfined phase, 
the universal exponents would be a key ingredient. 
It is also interesting to consider the thermalization process 
of the deconfined phase \cite{Hashimoto}. 

We believe that the Schwinger effect in the confining phase plays an important role  
in revealing new aspects of QCD with a strong electric field. 


We thank Y.~Ookouchi, G.~-W.~Semenoff, H.~Shimada, and F.~Sugino for useful discussions.

\end{document}